\documentclass{optica-article}

\journal{opticajournal} 

\articletype{Research Article}

\begin{document}

\title{Characterization of an $\rm ^{27}Al^+$  ion optical clock laser with three independent methods}

\author{Zhiyuan Wang, Zhiyu Ma, Wenzhe Wei, Jialu Chang, Jingxuan Zhang, Qiyue Wu, Wenhao Yuan, Ke Deng, Zehuang Lu, and Jie Zhang\authormark{*}}

\address{MOE Key Laboratory of Fundamental Physical Quantities Measurement\&Hubei Key Laboratory of Gravitation and Quantum Physics, PGMF and School of Physics, Huazhong University of Science and Technology, Wuhan 430074, P. R. China\\}

\email{\authormark{*}jie.zhang@mail.hust.edu.cn} 


\begin{abstract*} 
We report on the development and performance evaluation of an ultra-stable clock laser for an $\rm ^{27}Al^+$ optical clock. The thermal noise limited ultra-stable laser is developed based on a 30 cm long ultra-stable cavity. Three independent evaluation methods, including the frequency noise summation method, the three-cornered hat (TCH) method, and the optical clock transition detection method, are used to evaluate the clock laser performance. The summation result of various frequency noise terms is compared with the result of the TCH method. In addition, the $\rm ^{27}Al^+$ ion optical clock transition with ultra-narrow linewidth is also used to detect the frequency noise of the laser at lower Fourier frequencies. The results of the three methods show good agreements, showing a frequency instability level of $1.3\times10^{-16}$, and giving us confidence that these evaluation methods may provides guidance for accurate evaluations of high stability laser sources.
\end{abstract*}

\section{Introduction}
\label{sec:1}
Optical clocks based on narrow transitions of ions or neutral atoms expand the scope of scientific exploration due to their extremely low frequency instabilities \cite{safronova2018search,filzinger2023improved,su2018low}. A new definition of the ``second'' based on high accuracy comparisons of optical clocks is under consideration, which will bring revolutions to SI units definitions and the related precision measurement physics \cite{mcgrew2019towards}. Apart from time and frequency metrology, optical clocks can be used in fields like relativistic geodesy and fundamental physics \cite{grotti2018geodesy,safronova2018two}. A millimeter-level geoid height  difference measurement has been realized using a Sr optical lattice clock \cite{bothwell2022resolving}.

The performances of optical clocks are closely related to the frequency noises of ultra-stable clock lasers. The performances of clock lasers can directly affect the measured optical clock transition linewidths, or can prevent optical clocks reaching quantum projection noise (QPN) limit through Dick effect \cite{jiang2011making,schioppo2017ultrastable}. Accurate evaluations of laser frequency noises and frequency instabilities are essential to find out the main limits of the clock lasers. 

To evaluate the frequency noise of ultra-stable lasers accurately, the three-cornered-hat (TCH) method is commonly used \cite{matei20171,zhang2017ultrastable}. The TCH method relies on the availability of two more reference lasers with similar frequency instability levels and low correlations \cite{premoli1993revisited}. The TCH method is expensive and difficult to be implemented in a regular lab. An optical clock transition can also be used as an optical spectrum analyzer to evaluate the frequency noise of the clock laser \cite{bishof2013optical}. This evaluation method is convenient for ultra-stable lasers that are specifically developed as clock lasers, since no extra reference lasers are needed. The above two methods cannot give the information of the limiting noise contribution. Comprehensive noise evaluations are required to develop a high performance ultra-stable laser \cite{didier2019946,amairi2014long}. Here we demonstrate that, by directly summing all the different frequency noise contributions, the real frequency noise of the laser under evaluation can be predicted with the same level of accuracy. With this method, most of the frequency noise sensitivity coefficients can be measured through modulation experiments, thus lessen the requirements of the reference lasers. 

Among many optical clock transitions, the ultra-narrow $^1S_0\rightarrow\!^3P_0$ transition of $\rm ^{27}Al^+$ ion shows great potential for its insensitivity to blackbody radiation  \cite{chou2010frequency}. Since the clock transition wavelength is at deep UV of 267.4 nm, extremely low frequency noise is required for the clock laser. The frequency noises of ultra-stable lasers are ultimately limited by the thermal noises of the ultra-stable cavities \cite{numata2004thermal}. Various methods, including using longer cavity or working under cryogenic environment, have been used to reduce the thermal noise limit \cite{hafner20158,zhadnov202148,robinson2019crystalline}. While we have been working on clock laser development based on both methods \cite{zeng2018thermal,he2023ultra}, in this paper we focus on the system development and performance evaluations of an $\rm ^{27}Al^+$ ion optical clock laser based on a 30 cm long ultra-stable cavity. 

The main noises of the clock laser, including vibrational noise and temperature fluctuation noise, are evaluated and suppressed. Highly efficient methods of vibration sensitivity optimization and zero-crossing temperature measurement of the ultra-stable reference cavity are developed. After series of noise evaluations and optimizations, the total frequency noise is predicted by directly summing the different frequency noise contributions. The calculated laser performance with this method is compared to the evaluation results of the TCH method and the optical clock transition detection method. A thermal noise limited result is obtained with a frequency instability level of $1.3\times10^{-16}$. This result is confirmed by the all three evaluation methods. Benefits from the low frequency instability and the low long-term frequency drift of the ultra-stable clock laser, the linewidth of the $\rm ^{27}Al^+$ ion clock transition is reduced from 38(4) Hz to 2.9(5) Hz \cite{ma2020investigation}.

\section{Experimental setup}
\label{sec:2}
\begin{figure}[h]
\centering\includegraphics[width=0.7\textwidth]{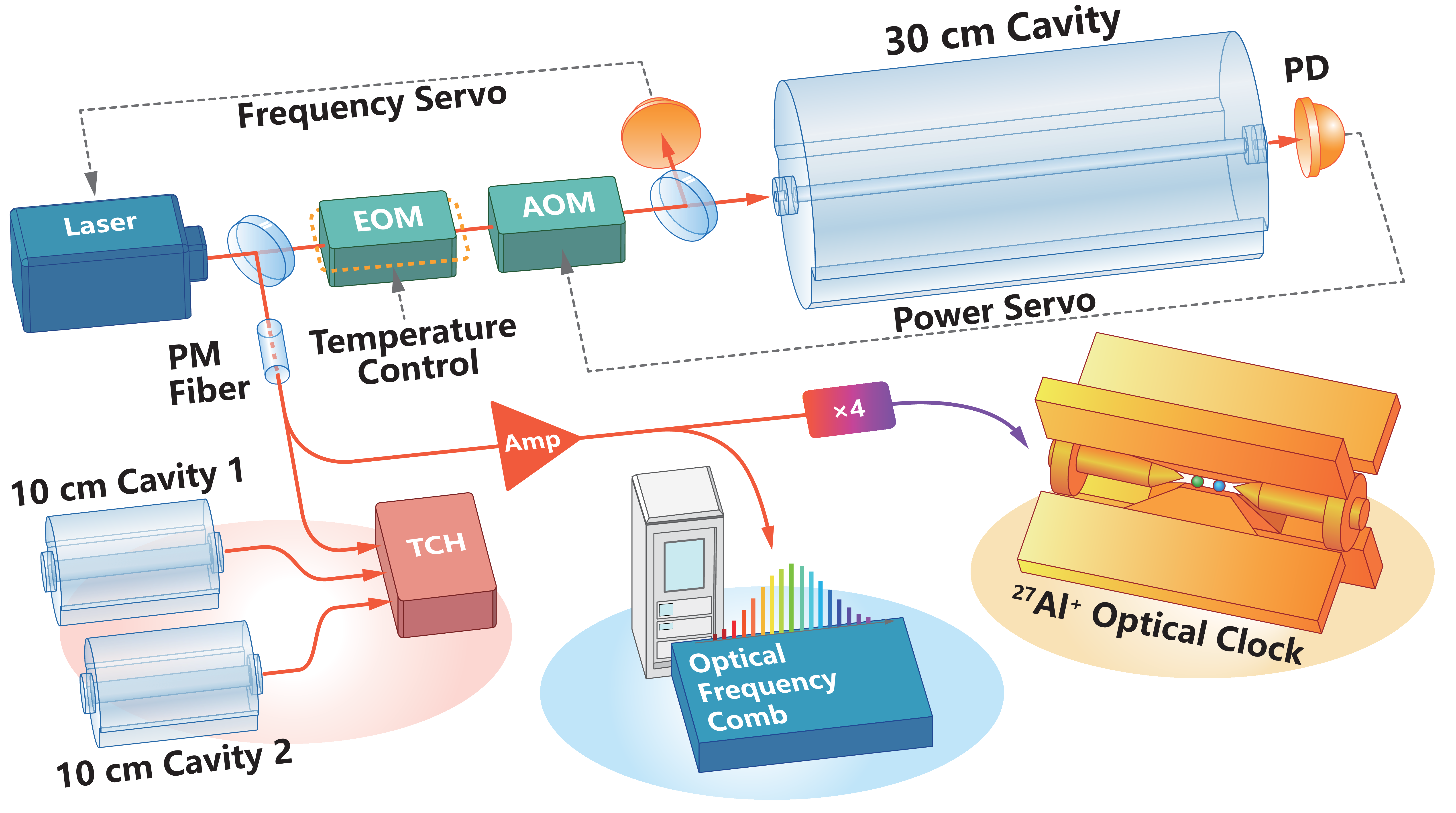}
\caption{\label{fig1} The frequency stabilization and performance evaluation setup of the 30 cm long cavity stabilized laser for $\rm ^{27}Al^+$ ion optical clock.}
\end{figure}

The experimental setup of the ultra-stable clock laser is shown in Fig. \ref{fig1}. A 1070 nm external cavity diode laser (Toptica, DL pro) is frequency stabilized to a 30 cm long ultra-stable cavity using the Pound-Drever-Hall (PDH) method. The cavity has a sandwich structure with spacer and compensation rings made of ultra low expansion (ULE) glass, while a pair of high reflectivity fused silica (FS) mirrors are attached to the spacer. The finesse of the cavity is measured to be 520,000 by the cavity ring-down (CRD) method, from which a cavity linewidth of 0.9 kHz can be inferred. We use an analog locking module (Toptica, FALC110) to servo control the laser frequency. A 2 MHz locking bandwidth is achieved after operational parameters optimization. The cavity is enclosed in a vacuum chamber with a pressure below $5\times10^{-6}$ Pa, and two layers of copper thermal shield are used inside the vacuum chamber to suppress the temperature fluctuation effect. Active temperature control are simultaneously performed on the outer copper thermal shield and the vacuum chamber. 

In the laser beam path, a resonant electro-optical modulator (EOM) is used to generate the phase modulation sidebands for PDH locking. An active temperature controlled thermal shield is used to reduce the temperature fluctuation of the EOM, thus the residual amplitude modulation (RAM) effect can be greatly suppressed. A photo detector (PD) is installed behind the cavity to monitor the laser power fluctuation, and is used for active laser power stabilization through an acoustic-optic modulator (AOM). The frequency stabilized laser is splitted and transmitted through a polarization-maintaining (PM) fiber for the TCH evaluation and the optical clock probing. The fiber induced noise is active compensated to an instability level below $10^{-17}$. At the optical clock lab, the 1070 nm ultra-stable laser is amplified and frequency quadrupled to 267.4 nm. One part of the amplified laser is transmitted to the optical frequency comb lab. The frequency comb is referenced to a hydrogen maser, and is used to measure the absolute frequency of the clock laser.

\section{Frequency noises evaluation and suppression}
\label{sec:3}
Precise frequency noises evaluation and suppression are indispensable for the development of high performance ultra-stable lasers. A detailed noise evaluation helps to predict the actual performance of the laser system. The major noise contribution terms in ultra-stable lasers have been widely discussed before \cite{webster2008thermal,keller2014simple,alvarez2019optical}. In general, the main frequency noise contributions are from vibration and temperature fluctuation of the environment, while further improvements depends on the suppression of noise terms like laser power jitters and RAM, etc. The ultimate performances of the lasers will be limited by the thermal noises of the ultra-stable cavities, which are the development goals of all the high performance ultra-stable lasers.

We calculate the thermal noise of the 30 cm long cavity by considering the Brownian noise and the thermo-elastic noise contributions. Other thermal noise sources like the thermo-refractive noise and the photo-thermo-elastic noise are estimated to have negligible contributions \cite{evans2008thermo,gorodetsky2008thermal}. According to the fluctuation-dissipation theorem, the Brownian noise from different parts of the cavity can be calculated based on the elastic strain energy under the action of an virtual force. The Brownian noises of the spacer, the mirror substrate, the mirror coating and the compensation ring can be calculated according to well developed theory \cite{kessler2012thermal,harry2002thermal}. We estimate the elastic strain energy of the large spacer of the 30 cm long cavity using finite element analysis (FEA) as the theoretical equations fail under such situation. The mirror coating contributes the most part in the Brownian thermal noise, which is calculated to be $4.0\times10^{-34}/f\;\rm m^2/Hz$.

Other than the Brownian noise, the coating thermo-elastic noise accounts for the main part at high Fourier frequency end of the thermal noise. The coating thermo-elastic noise of the 30 cm long cavity is calculated to be $5.9\times10^{-34}/\sqrt{f}\;\rm m^2/Hz$ \cite{braginsky2003thermodynamical}. 

\begin{figure}[h]
\centering\includegraphics[width=0.55\textwidth]{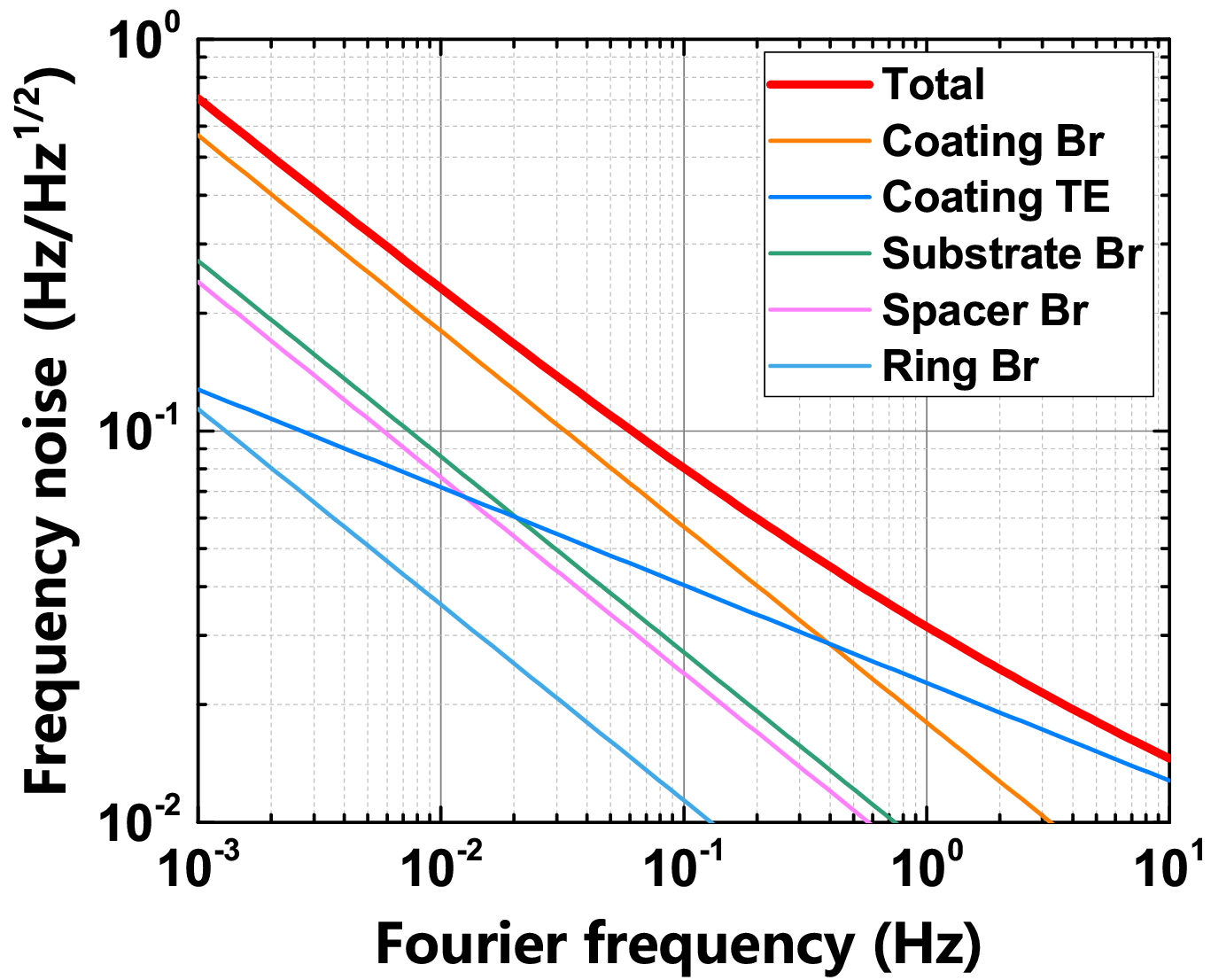}
\caption{\label{fig2} Thermal noise of the 30 cm long cavity. The total thermal noise is shown in red thick line and the contributions from different parts are shown in thin lines. Br: Brownian noise, TE: thermo-elastic noise.}
\end{figure}

The total thermal noise and contributions from different cavity parts are shown in Fig. \ref{fig2}. The FS mirrors with $\rm Ta_2O_5/SiO_2$ dielectric coatings contribute the main thermal noise. The total thermal noise contribution is calculated to be $0.032\;\rm Hz/\sqrt{Hz}@1\;\rm{Hz}$, which means that the frequency instability limit of the cavity stabilized laser is around $1.1\times10^{-16}@1\;\rm{s}$ when expressed with modified Allan deviation.

The development of thermal noise limited ultra-stable lasers depends on effective frequency noise suppression. The targets of the suppressed different frequency noise contributions are at least one order of magnitude lower than the cavity thermal noise limit. Because the vibrational noise and the temperature fluctuation noise are inevitable problems in high performance ultra-stable lasers using longer cavities or cryogenic environments, the development of efficient noise evaluation and suppression methods for these two effects have great values.

The effect of vibration on the laser frequency noise can be evaluated through measurement of the cavity vibrational sensitivity. For horizontal placed cylindrical cavity, the height and the transverse distribution of the mounting points are the key factors for reducing the vibrational sensitivity \cite{webster2007vibration,jin2018laser}. The preliminary choice is to support the cavity at its Airy points, and a further minor adjustment may be needed to achieve the lowest vibrational sensitivity. This adjustment needs repetitive mounting positions changes and vibrational sensitivity measurements. To avoid repeat vacuum pumping and waiting for temperature stabilization, we compare the vibrational sensitivities measured under vacuum and under atmospheric pressure by introducing artificial external vibrations. According to our evaluation, a vibrational sensitivity level no less than $10^{-10} /\rm g$ can be measured under atmospheric pressure without active temperature control.

In order to adjust the mounting positions of the cavity at precise steps, we design an adjustable mounting structure that can change the supporting positions using gaskets in a step of 1 mm. Teflon fixtures are also used during installation for fast cavity positioning. All of the methods mentioned above accelerate the optimization of the vibrational sensitivity. After adjusting the mounting positions of the cavity and replacing with a stable mounting structure, the three direction vibrational sensitivities decrease to $\kappa_x=89(2)\:\rm{kHz/g},\ \kappa_y=8(2)\:\rm{kHz/g},\ \kappa_z=113(1)\:\rm{kHz/g}$ from $\sim1\:\rm{MHz/g}$, where $x$ direction is the direction perpendicular to the cavity optical axis in horizontal plane, $y$ direction is the vertical direction, and $z$ direction is the cavity optical axis direction. This improvement of the vibrational sensitivity brings one order of magnitude better vibrational noise suppression. 

Considering the relatively large volume of the 30 cm long cavity, the cavity length change caused by temperature fluctuations and temperature inhomogeneity can be significant. Two key factors including the coefficient of the thermal expansion (CTE) and the temperature fluctuations of the cavity determine the influence of the environment temperature changes. Commonly used ULE cavities show extremely low CTE near their zero-crossing CTE temperatures. We develop a fast zero-thermal-expansion temperature measurement method by measuring a step temperature impulse response, which reduces the measurement time from several months to a few days for ultra-stable lasers with large thermal time constants. The details of this method are described in \cite{wang2021single}. Using different methods, the zero-crossing CTE temperature of the 30 cm long cavity is determined to be 4.3$\pm$0.5 $^{\circ}$C. Using two layers of active temperature control and with the help of passive thermal shields, the temperature fluctuation of the cavity is evaluated to be less than $\pm$0.6 $\rm\mu K$ in 12 hours. The 30 cm long cavity system is temperature controlled at 4.8 $^{\circ}$C in daily running and the contribution of the temperature noise is calculated to be less than 0.00005/$f\;\rm Hz/\sqrt{Hz}$ in frequency range higher than 1 mHz. The frequency drift of the 30 cm long cavity system is measured to be 10 mHz/s using an optical frequency comb that is referenced to a hydrogen maser. The drift rate is quite small among all the similar ULE cavities \cite{jiang2011making,sanjuan2019long,jin2018laser,didier2019946}, indicating good zero-crossing CTE temperature control.

\begin{figure}[h]
\centering\includegraphics[width=0.75\textwidth]{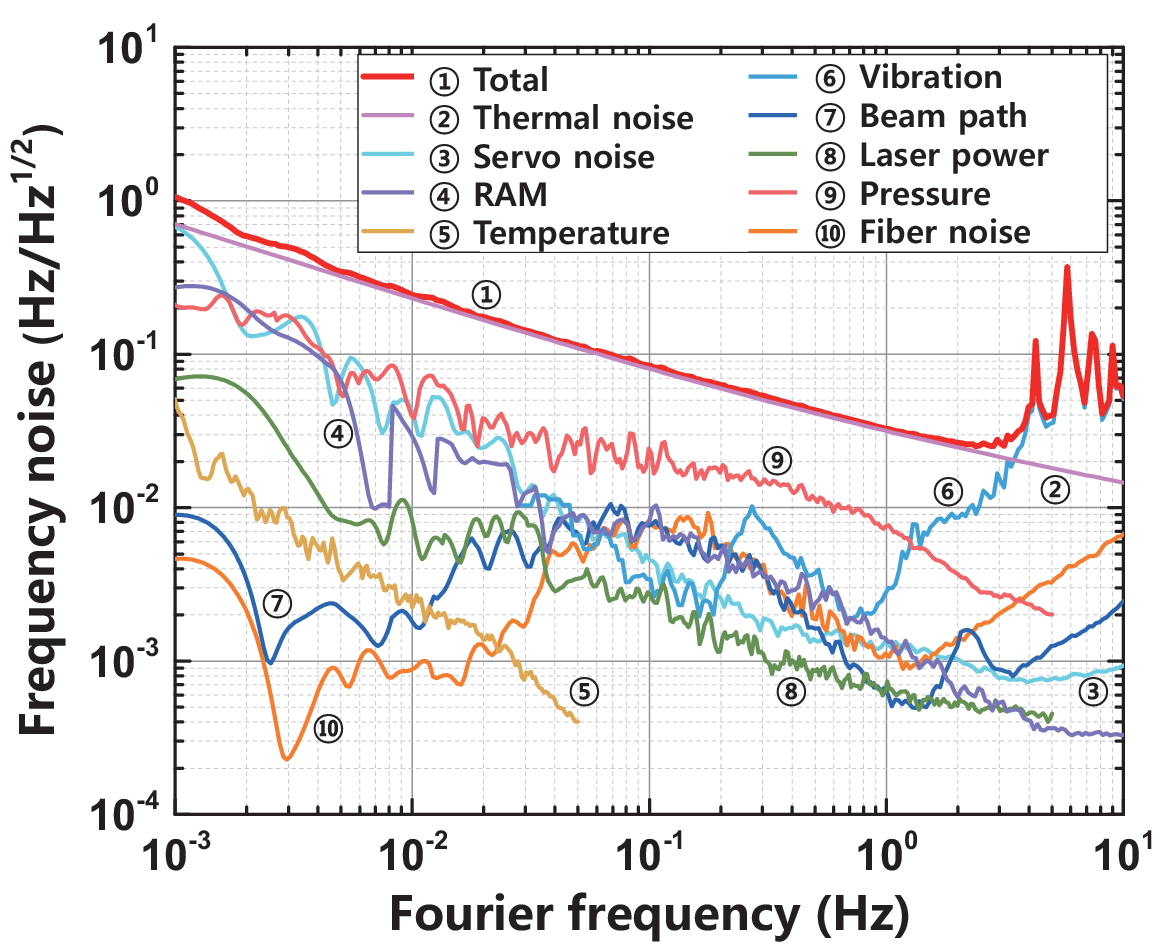}
\caption{\label{fig3} Laser frequency noise contributions. The red thick line represents the summation of the different noise terms that we have evaluated. The thin lines with different colors are the independent contributions of different noise terms.}
\end{figure}

In addition to the vibrational noise and the temperature fluctuation noise, other possible noise contributions are also carefully evaluated and suppressed, such as the laser power jitters, RAM, and the vacuum pressure fluctuations, etc. The frequency noise contributions from all these different noise sources are shown in Fig. \ref{fig3}, and all the noises are combined to predict the performance of the laser under the assumption that they are uncorrelated. Apart from the noises that we mentioned above, the servo noise is from the laser locking circuit, the beam path noise is the background noise of the laser beating beam path, and the pressure noise comes from the refractive index fluctuation inside the vacuum chamber. With all of the noises that we evaluated being suppressed below the thermal noise limit at Fourier frequency range of $0.01\sim1$ Hz, the total frequency noise shows a thermal noise limited result. At frequencies above 2 Hz the performance is limited by the vibrational noise.

\section{Performance evaluation using the TCH method}
\label{sec:4}
The TCH method, which is called the triangulation method originally, is introduced by Gray and Allan for frequency instability measurement of signal oscillators \cite{gray1974method}. The principle of this method is to compare the oscillator under test with two other reference oscillators. The TCH method has the advantage that the stability of the reference oscillators can be worse than the one under test. However, the correlations of the oscillators may result in a negative frequency instability result and the uncertainty of the measurement results will increase with worse references. Since the correlation of oscillators is hard to avoid in practice, a revised TCH method with the correlations being removed has been developed to solve this problem \cite{premoli1993revisited}. The algorithm of the TCH method is still under further optimization. A confidence interval estimation based on Bayesian statistics is developed to improve the reliability of the TCH method in long term frequency instability calculation \cite{vernotte2019confidence}.

To evaluate the frequency instability of the 30 cm long cavity stabilized laser, two similar systems based on 10 cm long cavities with performance around $5\times10^{-16}$ are used in the TCH setup \cite{zeng2018thermal}. The three ultra-stable laser beams are transmitted to a TCH evaluation platform through PM fibers. The fiber noise and the beating beam path noise are evaluated and are shown in Fig. \ref{fig3}. The phase deviations of the three beatnotes are recorded using a multichannel phase and frequency counter (K+K, FXE) for the TCH calculations.

\begin{figure}[h]%
\centering
\includegraphics[width=0.95\textwidth]{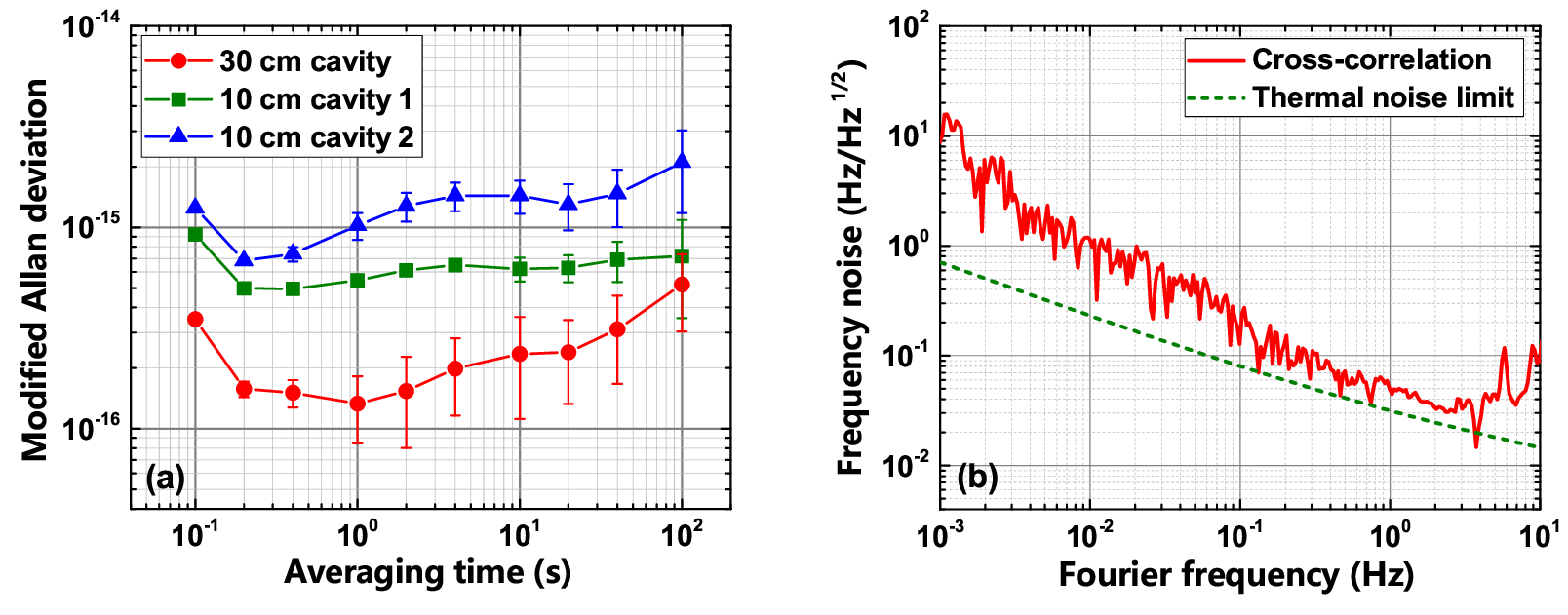}
\caption{(a) Measured frequency instabilities of the three ultra-stable lasers using the TCH method with the linear drifts removed. Error bars are the standard deviation of each point in 20 groups data statistics; (b) Calculated frequency noise of the 30 cm long cavity stabilized laser using cross-correlation method.}\label{fig4}
\end{figure}

Since the two 10 cm long cavity systems are in the same lab and share some common equipment, we use the revised TCH algorithm to remove the correlation. The frequency instabilities of the three ultra-stable lasers calculated using the TCH method are shown in Fig. \ref{fig4}(a). 20 groups of data each lasting 1000 seconds are used. From the evaluation result, the frequency instability of the 30 cm long cavity stabilized laser reaches $1.3(5)\times10^{-16}@1$ s in modified Allan deviation, and stays at $10^{-16}$ in the range from 0.1 s to 100 s. The frequency instabilities increases rapidly at the averaging time below 1 s. This is mainly caused by the high frequency vibrational noise, as we mentioned in Sec. \ref{sec:3}. Since the performances of the 10 cm long cavity systems are worse than that of the 30 cm cavity system, the calculated frequency instability of the 30 cm long cavity system shows bigger scatter, which confirms the peculiarity of the TCH method and can be seen in other similar TCH evaluations \cite{kessler2012sub}.

The frequency noise power spectral density (PSD) provides more detailed information of the ultra-stable laser system than that of Allan deviation. We measure the laser frequency noise using the cross-correlation method, which is widely used in microwave signal and laser noise analysis \cite{fest1983individual,xie2017phase}. The cross-correlation measurement setup is similar to the TCH method so we regard it also as the TCH evaluation here. The measured frequency noise of the 30 cm long cavity stabilized laser using the cross-correlation method is shown in Fig. \ref{fig4}(b). The frequency noise of the 30 cm long cavity system reaches the thermal noise limit at frequency range of $0.2\sim4$ Hz, and shows the same limitation of the high frequency vibrational noise as predicted by the previous noise evaluation method. The result of cross-correlation diverges from the previous section noise evaluation result at frequencies below 0.2 Hz. 

\begin{figure}[h]%
\centering
\includegraphics[width=0.95\textwidth]{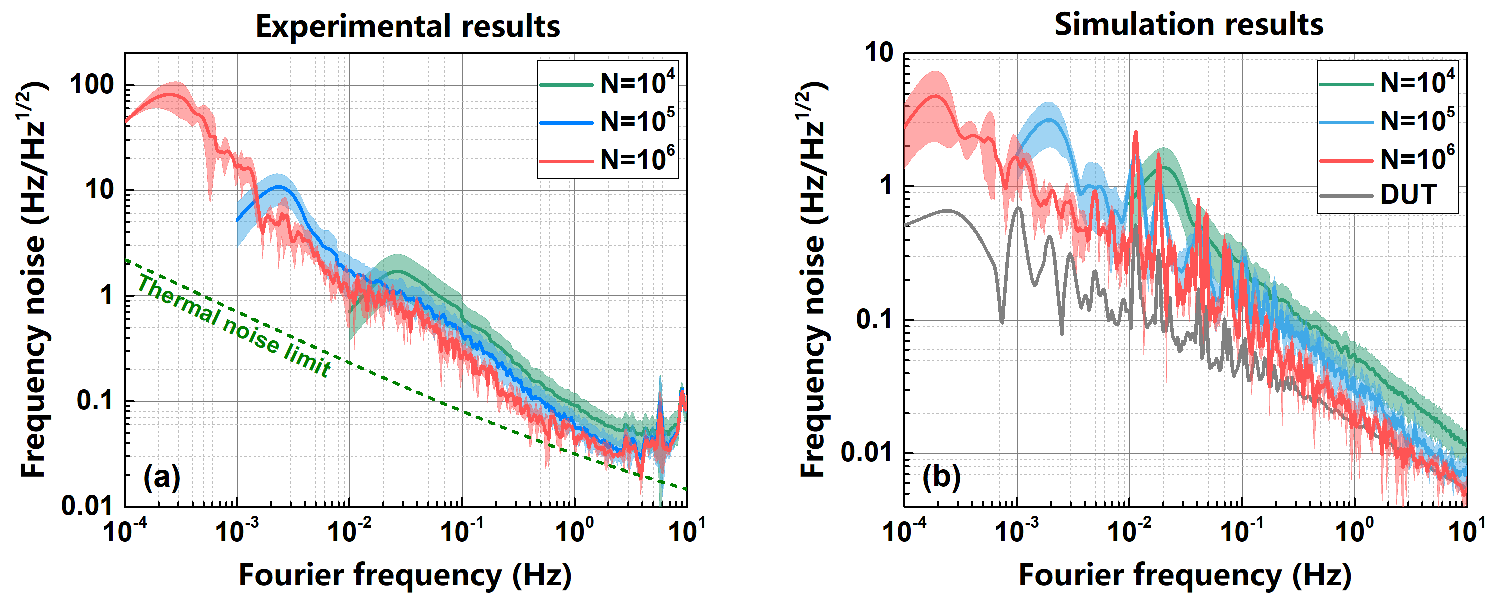}
\caption{(a) Cross-correlation results of the 30 cm cavity system under different sample number $N$. The solid lines with colored areas show the average and the standard deviation of $m$ groups linear PSD. The total sample number is $3\times10^{6}$, $m$=3, 30 and 300 for red, blue and green lines separately. The slope change at the lowest frequency of each line is due to the improved PSD algorithm we used \cite{trobs2006improved}. The green dashed line shows the thermal noise limit of the system; (b) Simulation results of cross-correlation method under different sample number $N$. The dark grey line gives the frequency noise of DUT, which is the target of cross-correlation calculation. 
}\label{fig5}
\end{figure}

This discrepancy can be understood by noticing that the calculated frequency noise of individual laser depends on the sample number $N$ for cross spectrum calculation, as shown in Fig. \ref{fig5}(a). The calculated frequency noise get closer to thermal noise limit of the 30 cm long cavity with the increasing of $N$. To avoid the result be influenced by some kinds of technical noises, we perform a simulation to confirm it and the simulation results are shown in Fig. \ref{fig5}(b). First, we generate three flicker noise series $x\rm{_A}(t)$, $x\rm{_B}(t)$ and $x\rm{_C}(t)$ to simulate thermal noises of the three cavities in our evaluation system. Then we produce the beat note $x\rm{_{AB}}(t)$ and $x\rm{_{AC}}(t)$. By calculating the cross spectrum of $x\rm{_{AB}}(t)$ and $x\rm{_{AC}}(t)$ we get the individual noise of the device under test (DUT) A, and it will be compared to directly calculated noise PSD of $x\rm{_A}(t)$. It can be seen in Fig. \ref{fig5}(b) that the results of cross-correlation converge to the noise of the DUT at frequency range higher than 1 Hz when $N=10^{6}$, it corresponds to at least $10^{4}$ segments for PSD calculation since the sample rate is 100. As is well-known in cross-correlation method, the cross-PSD will converge to noise PSD of the DUT when the average time $m\textgreater100$ under a white noise model \cite{rubiola2010cross}. Our simulation shows the segments number for PSD calculation needs to be larger than $10^{4}$ for $1/f$ noise model. Both the experimental and the simulation results in Fig. \ref{fig5} show the convergence of the cross-correlation method is not related to cross-PSD average times $m$ for $1/f$ noise model. The dependence to sample size or average times indicate the imperfection of the cross-correlation algorithm at the moment. And the actual situations may be more complicated when different noise models are mixed together.
  
\section{Performance evaluation using optical clock transition as probe}
\label{sec:5}
The frequency noise of the clock lasers influence the performance of optical clocks, which means that the optical clocks in return can be used to characterize the performance of the clock lasers. The $\rm ^{27}Al^+$ ion's environmentally insensitive, 8 mHz natural linewidth clock transition can serve as a perfect reference for clock laser frequency noise evaluation.

In an $\rm ^{27}Al^+$ ion optical clock, the quantum state of the ion may either in the ground state $\lvert g \rangle$ or in the excited state $\lvert e \rangle$ when it is probed with the clock laser. The frequency fluctuation of the clock laser causes the fluctuation of the state distribution, what can be described by population imbalance in a quantitative way. The Allan deviation of the population imbalance $\sigma_I$ can be expressed as:
\begin{equation}
\sigma_I=\int_{0}^{\infty}S_\nu\left(f\right)8\pi^2\sin^2\left(\pi fT_c\right)\lvert R\left(f\right)\rvert^2\mathrm{d}f
\label{eq1},
\end{equation}
where $S_\nu\left(f\right)$ is the frequency noise of the clock laser, $T_c$ is the cycle time of the clock, and $R\left(f\right)$ is the Fourier transform of the sensitivity function $g\left(t\right)$ when the ion is probed by the clock laser. The sensitivity functions show different spectral filtering properties depending on the spectroscopy probing sequences that are used to probe the clock transition. For example, the sensitivity functions of Rabi and Ramsey sequences show properties of low-pass filters, while sequences like photon echo can be used as a band-pass filter, and Unrig dynamical decoupling (UDD) sequence is a good high-pass filter \cite{bishof2013optical,cywinski2008enhance}. Using different timing sequences as filters, the frequency noise of the clock laser can be measured with higher signal to noise ratio (SNR). 

\begin{figure}[h]%
\centering
\includegraphics[width=0.8\textwidth]{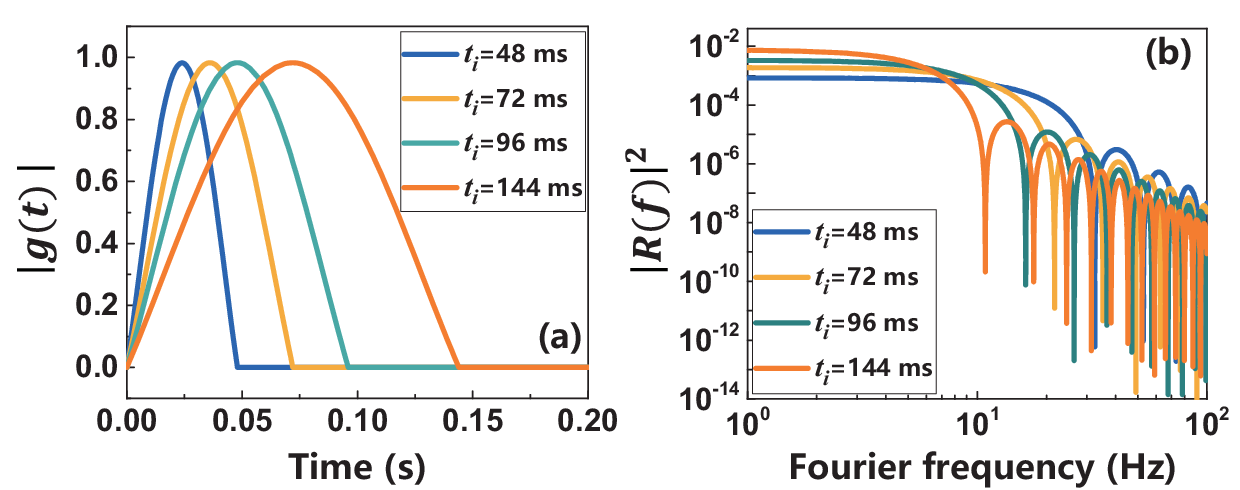}
\caption{Sensitivity functions of Rabi sequences with different probe times in (a) time domain; (b) frequency domain.}\label{fig6}
\end{figure}

The frequency noises of the ultra-stable clock lasers at low Fourier frequencies are more important to the performance of optical clocks. So we pay more attentions to the characterization of the low Fourier frequency end of the clock laser noise. To measure the low Fourier frequency end noise of the 30 cm long cavity laser system, we use Rabi sequences with different probe times $t_i$ to construct different low-pass filters. The sensitivity functions in time and frequency domains are shown in Fig. \ref{fig6}. Different probe times change the cut-off frequencies of the filters.

\begin{figure}[h]%
\centering
\includegraphics[width=0.9\textwidth]{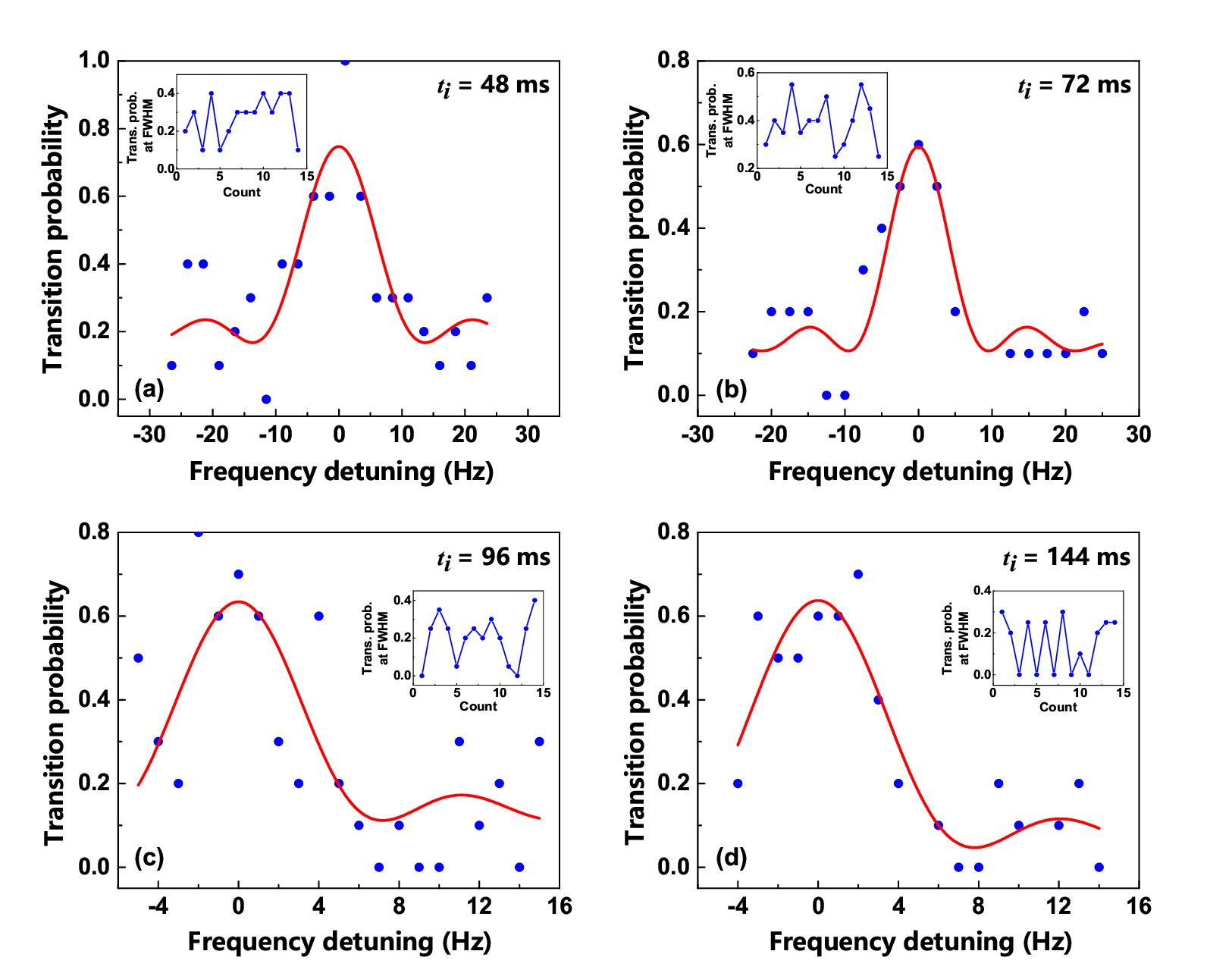}
\caption{The clock transitions and the fluctuations of the transition probabilities at the FWHM points (the inset figures) under different probe times. (a) $t_i$=48 ms. (b) $t_i$=72 ms. (c) $t_i$=96 ms. (d) $t_i$=144 ms.}\label{fig7}
\end{figure}

The measured clock transition probabilities under different probe times are shown in Fig. \ref{fig7}, each data point represents the average of 10 probes. The inset figures show the statistics of the transition probability at the full width at half maximum (FWHM) point of the clock transition line. The population imbalance $\sigma_I$ can be calculated by normalizing the Allan deviation of the transition probabilities to the maximum transition probability $P_{\rm max}$.

A laser noise model needs to be constructed before we estimate the frequency noise of the clock laser. After being filtered by the Rabi sensitivity function, the low Fourier frequency end laser noise can be obtained. Assuming the low Fourier frequency end laser noise is a combination of the white frequency noise and the thermal noise, then the frequency noise of the laser can be expressed as
\begin{equation}
S_\nu\left(f\right)=h_{\rm white}+\frac{h_{\rm thermal}}{f}
\label{eq2},
\end{equation}
where $h_{\rm white}$ and $h_{\rm thermal}$ are the coefficients of the white frequency noise and the thermal noise, respectively. Substituting Eq. \ref{eq2} into Eq. \ref{eq1}, we obtain
\begin{equation}
\begin{split}
\sigma_I=h_{\rm white}\int_{0}^{\infty}8\pi^2\sin^2\left(\pi fT_c\right)\lvert R\left(f\right)\rvert^2\mathrm{d}f+\\h_{\rm thermal}\int_{0}^{\infty}8\pi^2\frac{1}{f}\sin^2\left(\pi fT_c\right)\lvert R\left(f\right)\rvert^2\mathrm{d}f
\label{eq3}.
\end{split}
\end{equation}

Using symbols $a$ and $b$ to represent the integration terms of Eq. \ref{eq3}, then we have $\sigma_I=ah_{\rm white}+bh_{\rm thermal}$. The coefficients $h_{\rm white}$ and $h_{\rm thermal}$ can be obtained by least-squares fitting and then we can calculate the frequency noise of the clock laser. 

\begin{figure}[h]%
\centering
\includegraphics[width=0.6\textwidth]{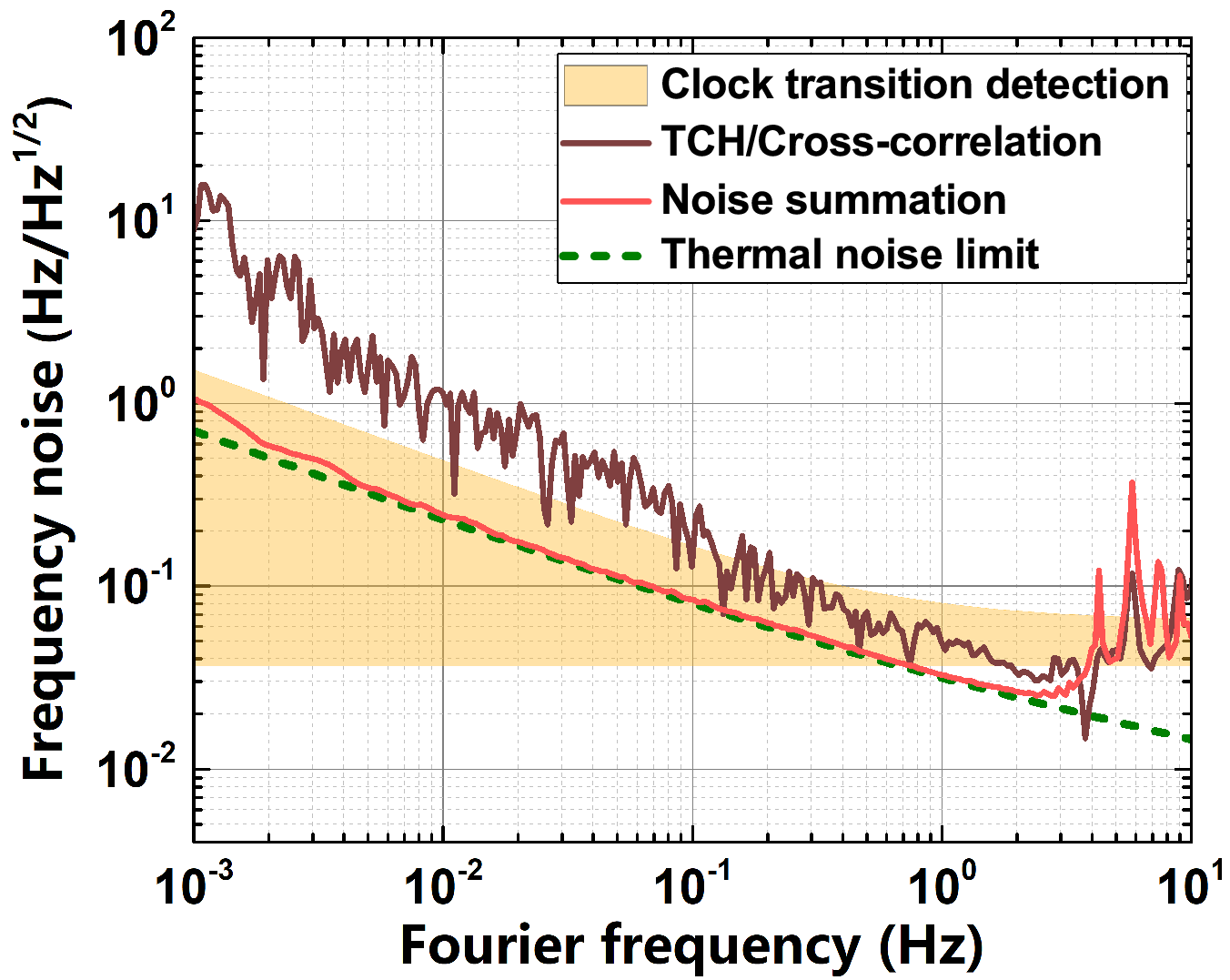}
\caption{Result comparison of the three evaluation methods we used. The orange area represents the laser frequency noise measured by the $\rm ^{27}Al^+$ ion optical clock transition. The frequency noise results of the cross-correlation and noise summation methods are shown in brown and red solid lines. The thermal noise limit is shown in green dashed line.}\label{fig8}
\end{figure}

The frequency noise of the 30 cm long cavity stabilized laser measured by the $\rm ^{27}Al^+$ ion optical clock transition is shown in Fig. \ref{fig8}. The orange area represents the distribution range of the measured laser frequency noise including the white noise and the thermal noise. It agrees well with the result of noise summation at low frequency range. It also agrees with the result of TCH/cross-correlation method at frequencies higher than 0.3 Hz, while it diverges from the TCH/cross-correlation result at lower frequency range due to limited sample size of the latter method. The evaluation results using optical clock depends on the noise model, which should be constructed in more detailed ways if a detailed full range noise is needed. The distribution range can be further reduced with better SNR of the clock transitions, which is related to the operation of the optical clocks. The comparison shows that the laser frequency noise prediction based on comprehensive noises evaluation can give accurate results without time and instrument costs of TCH/cross-correlation method or the usage of optical clocks. 

\begin{figure}[h]%
\centering
\includegraphics[width=0.57\textwidth]{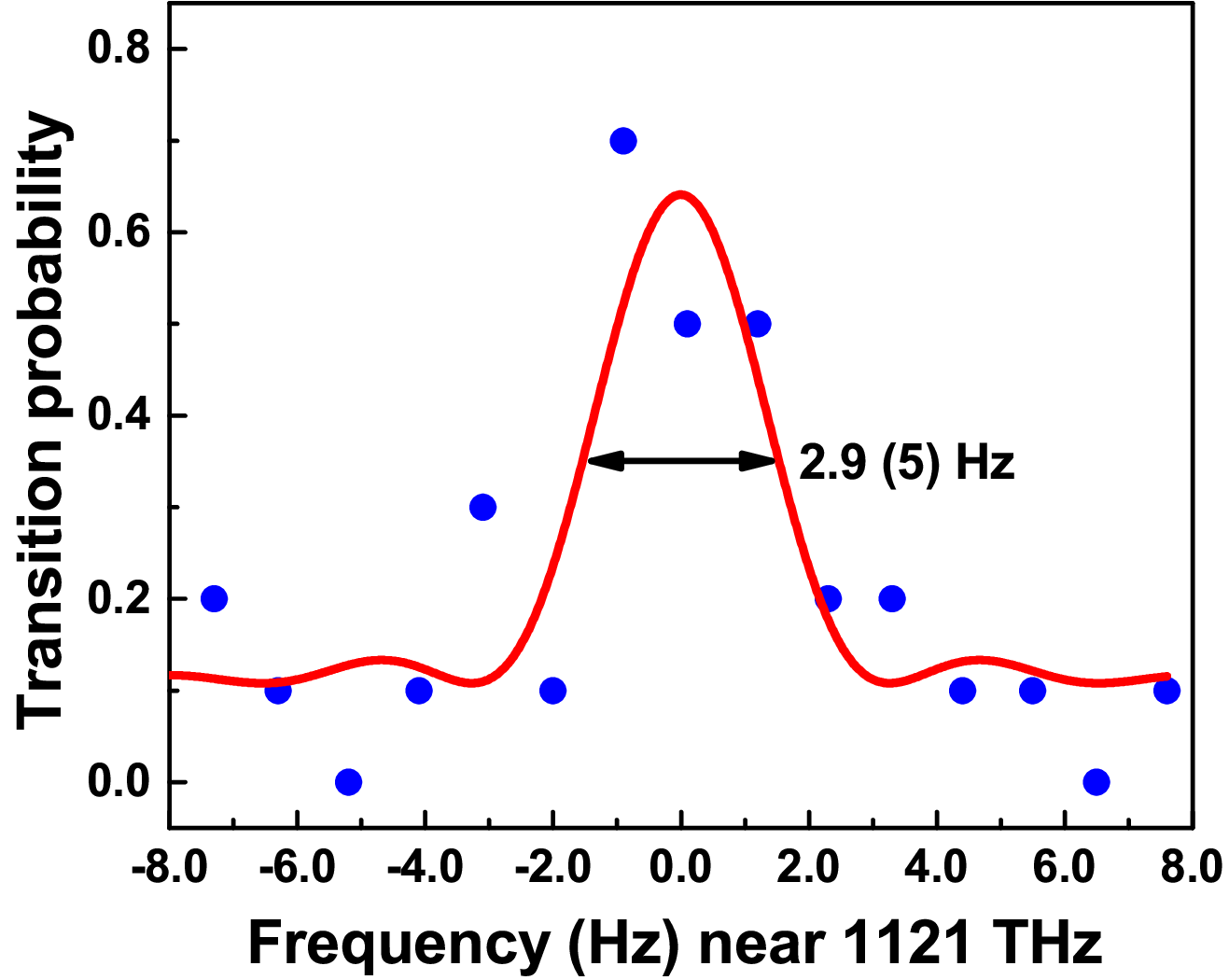}
\caption{Clock transition of the $\rm ^{27}Al^+$ ion optical clock using the 30 cm long cavity based clock laser.}\label{fig9}
\end{figure}

The 30 cm long cavity stabilized laser has been used as clock laser for the $\rm ^{27}Al^+$ ion optical clock. As shown in Fig. \ref{fig9}, an ultra-narrow linewidth clock transition of 2.9(5) Hz is detected due to the low frequency instability of the clock laser, what is comparable to the best result, as far as we know, to be reported on $\rm ^{27}Al^+$ ion optical clocks \cite{chou2010optical}.

\section{Conclusion}
\label{sec:6}
In conclusion, we have developed an ultra-stable clock laser for $\rm ^{27}Al^+$ ion optical clock with a frequency instability of $1.3\times10^{-16}$ based on a 30 cm long ultra-stable cavity, and have evaluated its performance using three independent methods. High-efficiency methods are developed to evaluate and suppress the vibrational noise and the temperature fluctuation noise of the ultra-stable clock laser. All the main noise contributions of the laser including the evaluation system noises are suppressed to smaller than the cavity's thermal noise limit. Direct summation of all the noise contributions is used to predict the frequency noise of the ultra-stable clock laser. The actual laser frequency noise is confirmed by using the THC/cross-correlation method and the $\rm ^{27}Al^+$ ion optical clock transition detection method. All three methods agree well at interested frequency range, while the cross-correlation measurement and optical clock detection show some differences due to their specific method related characteristics.

This work offer a general guidance for developing high-performance cavity stabilized lasers. The noise evaluation and suppression methods used here help to develop similar systems fast and systematically. The frequency noise evaluation using the $\rm ^{27}Al^+$ ion optical clock transition is the first time, as far as we know, to be realized on an ion optical clock. The laser noise evaluation using the optical clock transition has great values in better understanding the influence of the laser parameters to the optical clocks.

\begin{backmatter}
\bmsection{Funding}
This research was funded by Key Technologies Research and Development Program of China (2022YFC2204002, 2022YFB3904001); National Natural Science Foundation of China (61875065, 12004129, 62105109, 11904112); Special Project for Research and Development in Key areas of Guangdong Province (2019B030330001).
\bmsection{Acknowledgments}
We acknowledge fruitful discussions with R. Lalezari from FiveNine Optics concerning the high finesse mirrors.
\bmsection{Disclosures}
The authors declare no conflicts of interest.
\bmsection{Data availability}
Data underlying the results presented in this paper are not publicly available at this time but may be obtained from the authors upon reasonable request.
\end{backmatter}

\bibliography{references}

\begin{thebibliography}{10}
\newcommand{\enquote}[1]{``#1''}

\bibitem{safronova2018search}
M.~Safronova, D.~Budker, D.~DeMille, \emph{et~al.}, \enquote{Search for new
  physics with atoms and molecules,} {\protect\JournalTitle{Reviews of Modern
  Physics}} \textbf{90}, 025008 (2018).

\bibitem{filzinger2023improved}
M.~Filzinger, S.~D{\"o}rscher, R.~Lange, \emph{et~al.}, \enquote{Improved
  limits on the coupling of ultralight bosonic dark matter to photons from
  optical atomic clock comparisons,} {\protect\JournalTitle{Physical Review
  Letters}} \textbf{130}, 253001 (2023).

\bibitem{su2018low}
J.~Su, Q.~Wang, Q.~Wang, and P.~Jetzer, \enquote{Low-frequency gravitational
  wave detection via double optical clocks in space,}
  {\protect\JournalTitle{Classical and Quantum Grav.}} \textbf{35}, 085010
  (2018).

\bibitem{mcgrew2019towards}
W.~F. McGrew, X.~Zhang, H.~Leopardi, \emph{et~al.}, \enquote{Towards the
  optical second: verifying optical clocks at the {SI} limit,}
  {\protect\JournalTitle{Optica}} \textbf{6}, 448--454 (2019).

\bibitem{grotti2018geodesy}
J.~Grotti, S.~Koller, S.~Vogt, \emph{et~al.}, \enquote{Geodesy and metrology
  with a transportable optical clock,} {\protect\JournalTitle{Nature Physics}}
  \textbf{14}, 437--441 (2018).

\bibitem{safronova2018two}
M.~S. Safronova, S.~G. Porsev, C.~Sanner, and J.~Ye, \enquote{Two clock
  transitions in neutral {Yb} for the highest sensitivity to variations of the
  fine-structure constant,} {\protect\JournalTitle{Physical Review Letters}}
  \textbf{120}, 173001 (2018).

\bibitem{bothwell2022resolving}
T.~Bothwell, C.~J. Kennedy, A.~Aeppli, \emph{et~al.}, \enquote{Resolving the
  gravitational redshift across a millimetre-scale atomic sample,}
  {\protect\JournalTitle{Nature}} \textbf{602}, 420--424 (2022).

\bibitem{jiang2011making}
Y.~Jiang, A.~Ludlow, N.~D. Lemke, \emph{et~al.}, \enquote{Making optical atomic
  clocks more stable with 10$^{-16}$-level laser stabilization,}
  {\protect\JournalTitle{Nature Photonics}} \textbf{5}, 158--161 (2011).

\bibitem{schioppo2017ultrastable}
M.~Schioppo, R.~C. Brown, W.~F. McGrew, \emph{et~al.}, \enquote{Ultrastable
  optical clock with two cold-atom ensembles,} {\protect\JournalTitle{Nature
  Photonics}} \textbf{11}, 48--52 (2017).

\bibitem{matei20171}
D.~Matei, T.~Legero, S.~H{\"a}fner, \emph{et~al.}, \enquote{1.5 $\mu$ m lasers
  with sub-10 m{H}z linewidth,} {\protect\JournalTitle{Physical Review
  Letters}} \textbf{118}, 263202 (2017).

\bibitem{zhang2017ultrastable}
W.~Zhang, J.~Robinson, L.~Sonderhouse, \emph{et~al.}, \enquote{Ultrastable
  silicon cavity in a continuously operating closed-cycle cryostat at 4 {K},}
  {\protect\JournalTitle{Physical Review Letters}} \textbf{119}, 243601 (2017).

\bibitem{premoli1993revisited}
A.~Premoli and P.~Tavella, \enquote{A revisited three-cornered hat method for
  estimating frequency standard instability,} {\protect\JournalTitle{IEEE
  Transactions Instrum. and Meas.}} \textbf{42}, 7--13 (1993).

\bibitem{bishof2013optical}
M.~Bishof, X.~Zhang, M.~J. Martin, and J.~Ye, \enquote{Optical spectrum
  analyzer with quantum-limited noise floor,} {\protect\JournalTitle{Physical
  Review Letters}} \textbf{111}, 093604 (2013).

\bibitem{didier2019946}
A.~Didier, S.~Ignatovich, E.~Benkler, \emph{et~al.}, \enquote{946-nm {N}d:
  {YAG} digital-locked laser at 1.1$\times$10$^{-16}$ in 1 s and
  transfer-locked to a cryogenic silicon cavity,} {\protect\JournalTitle{Optics
  Letters}} \textbf{44}, 1781--1784 (2019).

\bibitem{amairi2014long}
S.~Amairi~ep Pyka, \enquote{A long optical cavity for sub-{Hertz} laser
  spectroscopy,} Ph.D. thesis (2014).

\bibitem{chou2010frequency}
C.~Chou, D.~Hume, J.~Koelemeij, \emph{et~al.}, \enquote{Frequency comparison of
  two high-accuracy {A}l$^+$ optical clocks,} {\protect\JournalTitle{Physical
  Review Letters}} \textbf{104}, 070802 (2010).

\bibitem{numata2004thermal}
K.~Numata, A.~Kemery, and J.~Camp, \enquote{Thermal-noise limit in the
  frequency stabilization of lasers with rigid cavities,}
  {\protect\JournalTitle{Physical Review Letters}} \textbf{93}, 250602 (2004).

\bibitem{hafner20158}
S.~H{\"a}fner, S.~Falke, C.~Grebing, \emph{et~al.},
  \enquote{8$\times$10$^{-17}$ fractional laser frequency instability with a
  long room-temperature cavity,} {\protect\JournalTitle{Optics Letters}}
  \textbf{40}, 2112--2115 (2015).

\bibitem{zhadnov202148}
N.~Zhadnov, K.~Kudeyarov, D.~Kryuchkov, \emph{et~al.}, \enquote{48-cm-long
  room-temperature cavities in vertical and horizontal orientations for {S}r
  optical clock,} {\protect\JournalTitle{Applied Optics}} \textbf{60},
  9151--9159 (2021).

\bibitem{robinson2019crystalline}
J.~M. Robinson, E.~Oelker, W.~R. Milner, \emph{et~al.}, \enquote{Crystalline
  optical cavity at 4 {K} with thermal-noise-limited instability and ultralow
  drift,} {\protect\JournalTitle{Optica}} \textbf{6}, 240--243 (2019).

\bibitem{zeng2018thermal}
X.~Zeng, Y.~Ye, X.~Shi, \emph{et~al.}, \enquote{Thermal-noise-limited
  higher-order mode locking of a reference cavity,}
  {\protect\JournalTitle{Optics Letters}} \textbf{43}, 1690--1693 (2018).

\bibitem{he2023ultra}
L.~He, J.~Zhang, Z.~Wang, \emph{et~al.}, \enquote{Ultra-stable cryogenic
  sapphire cavity laser with an instability reaching 2$\times$10$^{-16}$ based
  on a low vibration level cryostat,} {\protect\JournalTitle{Optics Letters}}
  \textbf{48}, 2519--2522 (2023).

\bibitem{ma2020investigation}
Z.~Ma, H.~Liu, W.~Wei, \emph{et~al.}, \enquote{Investigation of experimental
  issues concerning successful operation of quantum-logic-based $\rm
  ^{27}{A}l^+$ ion optical clock,} {\protect\JournalTitle{Applied Physics B}}
  \textbf{126}, 129 (2020).

\bibitem{webster2008thermal}
S.~Webster, M.~Oxborrow, S.~Pugla, \emph{et~al.},
  \enquote{Thermal-noise-limited optical cavity,}
  {\protect\JournalTitle{Physical Review A}} \textbf{77}, 033847 (2008).

\bibitem{keller2014simple}
J.~Keller, S.~Ignatovich, S.~A. Webster, and T.~E. Mehlst{\"a}ubler,
  \enquote{Simple vibration-insensitive cavity for laser stabilization at the
  10$^{-16}$ level,} {\protect\JournalTitle{Applied Physics B}} \textbf{116},
  203--210 (2014).

\bibitem{alvarez2019optical}
M.~D. {\'A}lvarez, \emph{Optical cavities for optical atomic clocks, atom
  interferometry and gravitational-wave detection} (Springer, 2019).

\bibitem{evans2008thermo}
M.~Evans, S.~Ballmer, M.~Fejer, \emph{et~al.}, \enquote{Thermo-optic noise in
  coated mirrors for high-precision optical measurements,}
  {\protect\JournalTitle{Physical Review D}} \textbf{78}, 102003 (2008).

\bibitem{gorodetsky2008thermal}
M.~L. Gorodetsky, \enquote{Thermal noises and noise compensation in
  high-reflection multilayer coating,} {\protect\JournalTitle{Physics Letters
  A}} \textbf{372}, 6813--6822 (2008).

\bibitem{kessler2012thermal}
T.~Kessler, T.~Legero, and U.~Sterr, \enquote{Thermal noise in optical cavities
  revisited,} {\protect\JournalTitle{Journal of the Optical Society of America
  B}} \textbf{29}, 178--184 (2012).

\bibitem{harry2002thermal}
G.~M. Harry, A.~M. Gretarsson, P.~R. Saulson, \emph{et~al.}, \enquote{Thermal
  noise in interferometric gravitational wave detectors due to dielectric
  optical coatings,} {\protect\JournalTitle{Classical and Quantum Grav.}}
  \textbf{19}, 897 (2002).

\bibitem{braginsky2003thermodynamical}
V.~B. Braginsky and S.~Vyatchanin, \enquote{Thermodynamical fluctuations in
  optical mirror coatings,} {\protect\JournalTitle{Physics Letters A}}
  \textbf{312}, 244--255 (2003).

\bibitem{webster2007vibration}
S.~Webster, M.~Oxborrow, and P.~Gill, \enquote{Vibration insensitive optical
  cavity,} {\protect\JournalTitle{Physical Review A}} \textbf{75}, 011801
  (2007).

\bibitem{jin2018laser}
L.~Jin, Y.~Jiang, Y.~Yao, \emph{et~al.}, \enquote{Laser frequency instability
  of 2$\times$10$^{-16}$ by stabilizing to 30-cm-long {F}abry-{P}{\'e}rot
  cavities at 578 nm,} {\protect\JournalTitle{Optics Express}} \textbf{26},
  18699--18707 (2018).

\bibitem{wang2021single}
Z.~Wang, Y.~Ye, J.~Chang, \emph{et~al.}, \enquote{Single step
  zero-thermal-expansion temperature measurement of optical reference
  cavities,} {\protect\JournalTitle{Optics Express}} \textbf{29}, 30567--30578
  (2021).

\bibitem{sanjuan2019long}
J.~Sanjuan, K.~Abich, M.~Gohlke, \emph{et~al.}, \enquote{Long-term stable
  optical cavity for special relativity tests in space,}
  {\protect\JournalTitle{Optics Express}} \textbf{27}, 36206--36220 (2019).

\bibitem{gray1974method}
J.~E. Gray and D.~W. Allan, \enquote{A method for estimating the frequency
  stability of an individual oscillator,} in \emph{28th annual symposium on
  frequency control,}  (IEEE, 1974), pp. 243--246.

\bibitem{vernotte2019confidence}
F.~Vernotte and {\'E}.~Lantz, \enquote{Confidence intervals for three-cornered
  hat and groslambert covariance estimates,} in \emph{2019 Joint Conference of
  the IEEE International Frequency Control Symposium and European Frequency and
  Time Forum (EFTF/IFCS),}  (IEEE, 2019), pp. 1--3.

\bibitem{kessler2012sub}
T.~Kessler, C.~Hagemann, C.~Grebing, \emph{et~al.}, \enquote{A
  sub-40-m{H}z-linewidth laser based on a silicon single-crystal optical
  cavity,} {\protect\JournalTitle{Nature Photonics}} \textbf{6}, 687--692
  (2012).

\bibitem{fest1983individual}
D.~Fest, J.~Groslambert, and J.~Gagnepain, Jean, \enquote{Individual
  characterization of an oscillator by means of cross-correlation or
  cross-variance method,} {\protect\JournalTitle{IEEE Transactions on
  Instrumentation and Measurement}} \textbf{32}, 447--450 (1983).

\bibitem{xie2017phase}
X.~Xie, R.~Bouchand, D.~Nicolodi, \emph{et~al.}, \enquote{Phase noise
  characterization of sub-hertz linewidth lasers via digital cross
  correlation,} {\protect\JournalTitle{Optics Letters}} \textbf{42}, 1217--1220
  (2017).

\bibitem{rubiola2010cross}
E.~Rubiola and F.~Vernotte, \enquote{The cross-spectrum experimental method,}
  {\protect\JournalTitle{arXiv preprint arXiv:1003.0113}}  (2010).

\bibitem{trobs2006improved}
M.~Tr{\"o}bs and G.~Heinzel, \enquote{Improved spectrum estimation from
  digitized time series on a logarithmic frequency axis,}
  {\protect\JournalTitle{Measurement}} \textbf{39}, 120--129 (2006).

\bibitem{cywinski2008enhance}
{\L}.~Cywi{\'n}ski, R.~M. Lutchyn, C.~P. Nave, and S.~D. Sarma, \enquote{How to
  enhance dephasing time in superconducting qubits,}
  {\protect\JournalTitle{Physical Review B}} \textbf{77}, 174509 (2008).

\bibitem{chou2010optical}
C.~Chou, D.~B. Hume, T.~Rosenband, and D.~J. Wineland, \enquote{Optical clocks
  and relativity,} {\protect\JournalTitle{Science}} \textbf{329}, 1630--1633
  (2010).

\end{thebibliography}






\end{document}